\documentclass[12pt,a4paper]{article}
\usepackage{amssymb}
\usepackage{epsfig}

\newcommand{\dfrac}[2]{{\displaystyle\frac{#1}{#2}}}
\newcommand{\be}{\begin{equation}}
\newcommand{\ee}{\end{equation}}

\begin{document}

\begin{center}
{\Large\bf Enhanced binding and cold compression of nuclei due to admixture of
antibaryons}\\[5mm]

{\bf T. B\"urvenich$^{\,1}$, I.N. Mishustin$^{\,1,2,3}$, 
L.M. Satarov$^{\,1,2}$,\\ 
J.A. Maruhn$^{\,1}$, H. St\"ocker$^{\,1}$, and W.~Greiner$^{\,1}$}
\end{center}
\begin{tabbing}
\hspace*{1.5cm}\=
${}^1$\={\it Institut~f\"{u}r~Theoretische~Physik,
J.W.~Goethe~Universit\"{a}t,}\\
\>\>{\it D--60054~Frankfurt~am~Main,~\mbox{Germany}}\\
\>${}^2$\>{\it The Kurchatov~Institute, Russian Research Center,}\\
\>\>{\it 123182~Moscow,~\mbox{Russia}}\\
\>${}^3$\>{\it The Niels~Bohr~Institute,
DK--2100~Copenhagen {\O},~\mbox{Denmark}}\\
\end{tabbing}
\vspace*{-5mm}
\begin{center}\today\end{center}

\begin{abstract}  
\noindent 
We discuss the possibility of producing a new kind of 
nuclear system by putting a few antibaryons inside ordinary nuclei.
The structure of such systems is calculated within the 
relativistic mean--field model assuming that the nucleon
and antinucleon potentials are related by the 
G--parity transformation. The presence of antinucleons leads to
decreasing vector potential and increasing
scalar potential for the nucleons.
As a result, a strongly bound system of high
density is formed. Due to the significant
reduction of the available phase space  the annihilation probability might
be  strongly suppressed in such systems. 
\end{abstract}

\section{Introduction}
Presently it is widely accepted that the relativistic
mean--field (RMF) model~\cite{RMF} gives a good description 
of nuclear matter and finite
nuclei~\cite{Rei89}. Within this approach the nucleons are supposed to
obey  the Dirac equation coupled to mean meson fields. Large 
scalar and vector potentials, of the order of 300 MeV, are necessary
to explain the strong spin--orbit splitting in nuclei. The most debated
aspect of this model is related to the negative--energy states of the
Dirac equation. In most applications these states are simply ignored
(no--sea approximation) or ''taken into account'' via the non--linear
and derivative terms of the scalar potential. On the other hand, explicit
consideration of the Dirac sea combined with the \mbox{G--parity}
arguments leads to such interesting conjectures as the existence of
deeply--bound antinucleon  states in nuclei~\cite{Aue} or even
spontaneous production of nucleon--antinucleon
pairs~\cite{Mis90,Mis93}. These
predictions are based on the assumption that the relativistic
description of the nucleon within the RMF model is a valid concept
(for a discussion see Refs.~\cite{Bro,Mos,Fur}).
Unfortunately, the experimental information on the antinucleon effective
potential in nuclei is obscured by the strong absorption caused by  
annihilation. As follows from the analysis of Ref.~\cite{Gal}, 
the real part of the antiproton effective potential might
be as large as 200--300 MeV, with the uncertainty reaching 100\% in the
deep interior of the nucleus.

Keeping in mind all possible limitations of the RMF approach, below we
consider yet another interesting application of this model. Namely, we
study properties of light nuclear systems containing a few real
antibaryons. At first sight this may appear ridiculous because
of the fast annihilation of antibaryons in the dense baryonic
environment. But as our estimates show, due to a significant reduction
of the available phase space for annihilation, the life time of such
states might be long enough for their observation. In a certain
sense, these states are analogous to the famous baryonium states in the
$N\overline{N}$ system \cite{Sha}, although their existence has never
been unambiguously confirmed. 

It should be emphasized that previous discussions of the
negative--energy states in nuclei were somewhat academic because
their contributions to the source terms  
for the meson fields were ignored. In contrast, our primary goal here 
is to study properties of
physical systems containing both real baryons and real antibaryons. 
There  could be different theoretical schemes for solving this problem 
but we find the RMF model most suitable for this study. To our
knowledge,  up till now a self--consistent calculation of 
antinucleon states in nuclei has not been performed. Our calculations
can be regarded as the first attempt to fill this gap. We consider two
nuclear systems, namely $^{16}$O and $^{8}$Be, and study the changes in
their structure due to the presence of an antiproton.

\section{Theoretical framework}
Below we use the RMF model which previously
has been successfully applied for describing ground--states of nuclei
at and away from the $\beta$--stability line. For nucleons, the scalar
and vector potentials contribute with opposite signs in the central
potential, while their sum enters in the spin--orbit potential. Due to
G--parity, for antiprotons the vector potential changes sign and
therefore both the  scalar and the vector mesons generate attractive
potentials. 

To estimate uncertainties of this approach we use three different
parametrizations of the model, namely NL3~\cite{Lal97},
NL--Z2~\cite{NLZ2} and TM1~\cite{TM1}. Their parameters are
found by fitting  binding energies and observables 
related to formfactors of spherical
nuclei from $^{16}$O (not included in the TM1 fit) to Lead isotopes. 
In NL3, properties of symmetric nuclear matter have been included 
in the fit as well. The TM1 model, implementing a
self--interaction of the $\omega$--meson, gives a softer rise of the
vector potential with density. This leads to smaller meson fields as
compared to the NL3 and NL--Z2 parametrizations. In this paper we assume that the
antiproton interactions are fully determined by the G--parity
transformation. 
\begin{figure*}[t!]
%\hspace*{-1cm}
\epsfig{figure=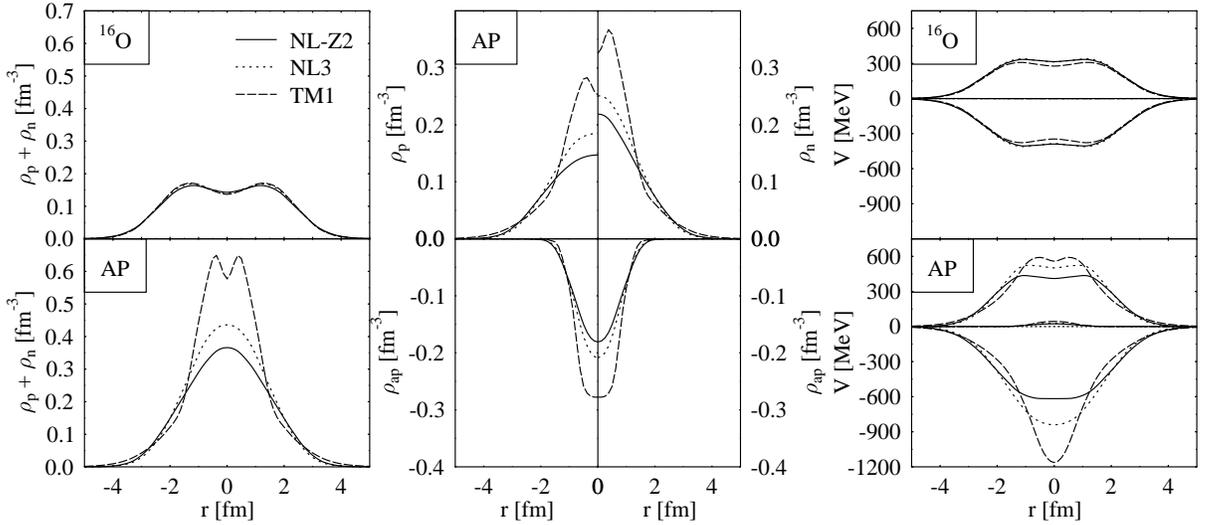,width=16cm}
\caption{\label{fig:o16_1}
The left panel represents the sum of proton and neutron densities
as function of nuclear radius for $^{16}$O without (top) and with
an antiproton (denoted by AP).  The left and right parts of the upper middle
panel show separately the proton and neutron densities, the lower
part of this panel displays the antiproton density.  The right panel
shows the scalar (negative)  and vector (positive) parts of the nucleon
potential.  Small contributions shown in the lower row correspond  to the
isovector ($\rho$--meson) part.} 
\end{figure*}
Following Mao et al. \cite{Mao99}, we solve  the effective
Schr{\"o}dinger equations for both the nucleons and the
antiprotons. Although we neglect the Dirac sea
polarization, we take into account explicitly
the contribution of the antibaryon into the scalar and
vector densities. 
For protons and neutrons  we include pairing correlations
within the BCS model with a \mbox{$\delta$--force} (volume
pairing)~\cite{Ben00}. Calculations are done within the blocking
approximation \cite{Rutz} for the antiproton, and assuming 
the time--reversal invariance of the nuclear ground--state. 
The energy of the system is found by using the damped gradient iteration
method~\cite{Rei82}. The coupled set of equations for nucleons,
antinucleons and mesons is solved iteratively and self--consistently.
The numerical code
employs axial and reflection symmetry, allowing for axially symmetric
deformations of the system.

\section{Structure of nuclei containing antiprotons}
As an example, we consider the nucleus $^{16}$O  with one antiproton in
the lowest bound state. This nucleus  is the lightest nucleus for which
the mean--field  approximation is acceptable, and it is included into
the fit of the effective forces NL3 and NL-Z2. The antiproton state is assumed
to be in the
$s1/2^+$ state. The antiproton contributes with the same sign as
nucleons to the scalar density, but with opposite sign to the vector
density. This leads to an overall increase of attraction and decrease
of repulsion for all nucleons. The antiproton becomes very deeply bound
in the $s1/2^+$ state. To maximize attraction, protons and neutrons 
move to the center of the nucleus, where the antiproton has its largest
occupation probability. This leads to a cold compression  of the
nucleus to a high density.  

Figure~\ref{fig:o16_1} shows the densities and potentials for $^{16}$O
with and without the antiproton.  For normal $^{16}$O all RMF
parametrizations considered produce very  similar results. The presence
of an antiproton  dramatically changes the structure of the nucleus. The
sum of proton and neutron densities reaches a  maximum value
of $(2-4)\,\rho_0$\,, where $\rho_0\simeq 0.15$\,fm$^{-3}$ is the normal
nuclear density, depending on the parametrization. The
largest compression is predicted by the TM1 model. This follows from
the fact that this parametrization gives the softest equation of state
as compared to other forces considered here. 

According to our calculations, the difference between proton and
neutron  densities is quite large, which leads to an increase in symmetry
energy.  The reason is that protons, though they feel additional
Coulomb attraction to the antiproton, repel each other. As a 
consequence, neutrons are concentrated closer to the center  than
protons and the symmetry energy increases.
\begin{figure*}[t!]
%\hspace*{-5mm}
\epsfig{figure=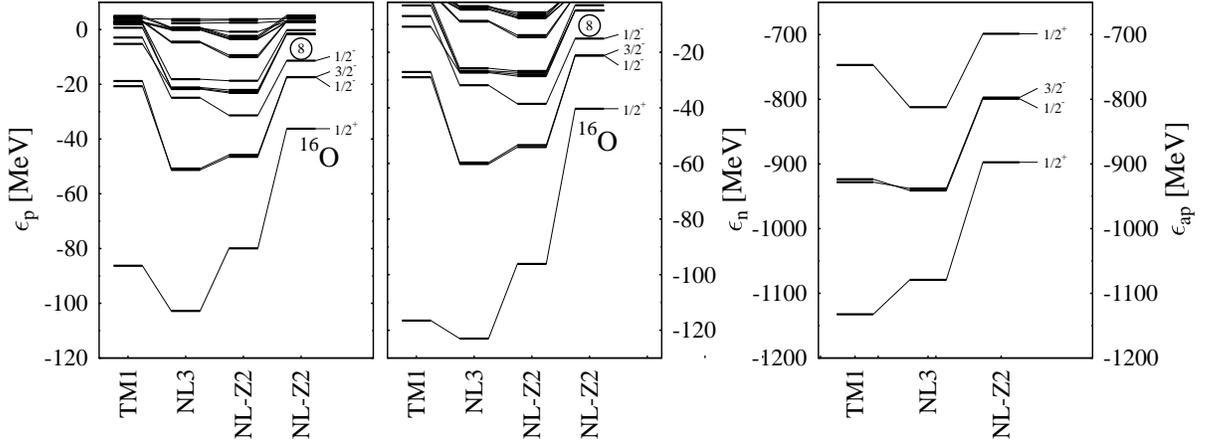,width=16cm}
\caption{\label{fig:o16_spec}
Proton (left), neutron (middle) and antiproton (right)
energy levels for the nucleus $^{16}$O with one antiproton and 
without it (rightmost columns in left and middle panel).
}
\end{figure*}
As compared to the case without the antiproton, absolute values of vector
and scalar potentials increase  in the central region of the nucleus.
This leads to an enormous  drop of the effective baryon mass near the
nuclear center, which should  strongly suppress local rates of
annihilation (see below).  The effective mass for TM1 even becomes 
negative at \mbox{$r\lesssim 1$\,fm}.

Since nucleons feel a deeper potential as compared to the nucleus
without the antiproton, their binding energy increases too. This can be 
seen in Fig.~\ref{fig:o16_spec}. The nucleon binding is largest within the
NL3 parametrization. In the TM1 case, the $s1/2^+$ state is
also deep, but higher levels are less bound as compared to the 
NL3 and NL--Z2 calculations. This is a consequence of the smaller
spatial extension of the potential in this case. The highest $s1/2^-$ level 
is even less bound than for the system without an antiproton.

For the antiproton levels, the TM1 parametrization predicts the 
deepest bound state with
binding energy of about $1130$~MeV. The NL3 calculation gives nearly
the same binding,  while in the NL--Z2 case, antiproton states are 
more shallow and have smaller spacing. It should be noted that the antiprotons 
are more strongly bound than was obtained in Ref.~\cite{Mao99}. This follows
from the fact that here we calculate both nucleons and the antinucleon 
self--consistently allowing the target nucleus to change its shape and 
structure due to the presence of the antiproton.
The total binding energy of the system is predicted to be~828~MeV for
NL--Z2, 1051~MeV for NL3, and 1159~MeV for TM1. 
For comparison, the binding energy of a normal $^{16}$O nucleus
is 127.8, 128.7 and 130.3 MeV in the case of NL--Z2, NL3, and
TM1, respectively.
Due to this anomalous binding we call these systems 
Super Bound Nuclei (SBN).

\begin{figure}[t]
\epsfig{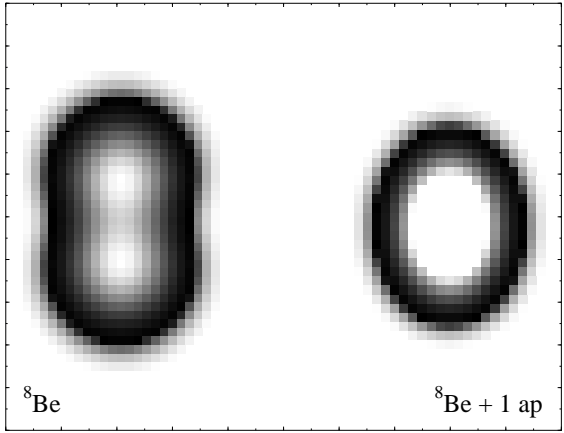}
\caption{\label{fig:be8}
Contour plot of nucleon densities for $^8$Be without (left) and
with (right) antiproton calculated with
the parametrization NL3. The maximum density of normal $^8$Be 
is $0.20~{\rm fm}^{-3}$, while for nucleus 
with antiproton it is $0.61~{\rm fm}^{-3}$.
}
\end{figure}
As a second example, we investigate the effect of a single antiproton
inserted into the $^8$Be nucleus. In this calculation only the NL3
parametrization was used (the effect is similar for all three forces). The normal $^8$Be  nucleus is not spherical,
exhibiting a clearly visible  $2~\alpha$~structure  with the
deformation $\beta_2\simeq 1.20$ in the ground--state.  Inserting the
antiproton gives rise to compression and change of nuclear shape,
resulting in a much less elongated  nucleus with  $\beta_2\simeq 0.23$.
Its maximum density increases by a factor of three from  $0.20~{\rm
fm}^{-3}$ to  $0.61~{\rm fm}^{-3}$. The cluster structure of ground
state completely vanishes. A similar effect has been predicted in
Ref.~\cite{KBe} for the case of the $K^-$ bound state  in the $^8$Be
nucleus. In the considered case the binding
energy increases  from $52.9$~MeV (the experimental value is 
$56.5$~MeV) to about $700$~MeV! 

\section{Life time, formation probability
and signatures of SBNs}
The crucial question concerning a possible observation of the SBNs 
is their life time. The only decay 
channel for such states is the annihilation on surrounding nucleons.
The mean life time of an antiproton in nucleonic matter of density
$\rho_B$ can be estimated as 
$\tau=<\sigma_Av_{\rm rel}\rho_B>^{-1}$\,, where 
angular brackets denote averaging over the
wave function of the antiproton and $v_{rel}$ is its relative  velocity
with respect to nucleons. In vacuum the $N\overline{N}$ annihilation
cross section at low $v_{\rm rel}$ can be parametrized as~\cite{Dov}
$\sigma_A=C + D/v_{\rm rel}$ with C=38 mb and D=35 mb. For 
$<\rho_B>\simeq 2 \rho_0$  this would lead to a very short  life time,
$\tau\simeq 0.7$ fm/c (for $v_{\rm rel}\simeq 0.2$). However, one
should bear in mind that the annihilation process is very sensitive to
the phase space available for decay products. For a bound nucleon and
antinucleon the available energy is $Q=2m_N-B_N-B_{\overline{N}}$,
where $B_N$ and $B_{\overline{N}}$ are the corresponding binding energies.
As follows from the calculations presented above
(see~Fig.~\ref{fig:o16_spec}), this  energy is strongly reduced
compared to $2m_N$, namely, $Q\simeq 600-680$ MeV (TM1), 810--880 MeV
(NL3) and 990--1050 MeV  (NL--Z2) for the lowest antiproton states.

For such low values of $Q$ many important annihilation channels
involving two heavy mesons ($\rho$, $\omega$, $\eta$, $\eta^\prime$, ...)  
are simply closed. Other two--body channels such as $\pi\rho$,
$\pi\omega$ are considerably suppressed due to the closeness to the
threshold. As is well known, the two--pion final states
contribute only about 0.4\% of the annihilation cross
section. Even in vacuum  all above mentioned  channels
contribute to $\sigma_A$ not more than 15\%~\cite{Ams}. Therefore, we
expect that only multi--pion final states contribute significantly to
antiproton annihilation in the SBN. But these channels 
are strongly suppressed
due to the reduction of the available phase space. Our calculations
show that changing $Q$ from 2 GeV to 1~GeV results in 
suppression factors 5, 40 and 1000 for the annihilation  channels with
3, 4 and 5 pions in the final state, respectively. Applying these suppression factors
to the experimental branching ratios \cite{Sed} we come to the
conclusion that in the SBNs the annihilation rates can be easily suppressed
by factor of 20--30. There could be additional suppression factors of
a structural origin which are difficult to estimate at present. This
brings the SBN life time to the level of \mbox{15--20 fm/c} which
makes their experimental observation feasible. The corresponding width,
$\Gamma\simeq 10$ MeV, is comparable to that of the $\omega$--meson.

Let us discuss now how these exotic nuclear states can be 
produced in the laboratory. We
believe that the most direct way is to use antiproton beams of
multi--GeV energy. This high energy is needed to suppress
annihilation on the nuclear surface which dominates at low
energies. To form a deeply bound state, the fast
antiproton must transfer its energy and momentum to one of the
surrounding nucleons. This can be achieved through
reactions of the type $\overline{p}p \rightarrow B\overline{B}$ in the nucleus,
\begin{eqnarray}
\overline{p} + (A,Z)\rightarrow B + {}_{\overline{B}\,}(A-1,Z^\prime)~,
\end{eqnarray}
where $B=n,p,\Lambda,\Sigma$. The fast antibaryon $B$ can be used as a trigger
of events where the antibaryon $\overline{B}$ is trapped in the nucleus. 
Of course, some additional soft pions can be emitted too.
One can think even
about producing an additional baryon-antibaryon pair and forming a
nucleus with two antibaryons in the deeply bound states. In this case two
fast nucleons will be knocked out from the nucleus. 

Without detailed transport calculations it is difficult  to
find the formation probability, $W$\,, of final nuclei with trapped 
antinucleons in these reactions. A rough estimate can be 
obtained by assuming that antiproton stopping is achieved in a
single inelastic collision somewhere in the nuclear
interior i.e. taking the penetration length 
of the order of the nuclear radius $R$\,.
From the Poisson distribution in the number of collisions
the probability of such an event~is
\be\label{pr1}
w_1=\dfrac{R}{\lambda_{\rm in}}\exp{\left(-\dfrac{R}{\lambda}\right)}\,,
\ee
where $\lambda_{\rm in}^{-1}=\sigma_{\rm in}\rho_0$ and
$\lambda^{-1}=(\sigma_{\rm in}+\sigma_A)\rho_0$ (here
$\sigma_{\rm in}$ and $\sigma_A$ are the inelastic and
annihilation parts of the $\overline{p}N$ cross section).
The exponential factor in Eq.~(\ref{pr1}) includes the
probability to avoid annihilation. For initial
antiproton momenta $p_{\rm lab}\simeq 10$ GeV 
we use $\sigma_{\rm in}\simeq 25$ mb,  
$\sigma_A\simeq 15$ mb~\cite{Sed} and get  
$\lambda\simeq 1.6$ fm which is comparable with the radii
of light nuclei. For an oxygen target, using $R\simeq 3$ fm
leads to $w_1\simeq 0.17$\,.   

In fact we need relatively small final antiproton momenta to overlap
significantly with the momentum distribution of 
a bound state, namely, $\Delta p\sim\pi/R_{\overline{p}}$\,, 
where $R_{\overline{p}}\simeq 1.5$ fm is characteristic 
size of antiproton spatial distribution (see Fig.~\ref{fig:o16_1}).
The probability of such a momentum loss can be estimated  
by the method of Refs.~\cite{Hwa,Cse} which was previously
used for calculating proton spectra in high--energy pA collisions. 
At relativistic bombarding energies the differential cross
sections of the $\overline{p}p\to\overline{p}X$ and $pp\to pX$
reactions are similar. The inelastic parts of these cross sections 
drop rapidly with transverse momentum, but they are practically flat as a
function of longitudinal momentum of secondary particles.
Thus, the probability of final antiproton momentum to fall in the interval 
$\Delta p$ is simply $\Delta p/p_{\rm lab}$\,.  
For $p_{\rm lab}=10$\,GeV and $\Delta p=0.4$\,GeV this gives 0.04. 
Assuming the geometrical fraction of central events $\sim 20\%$ we get  
the final estimate 
$W\simeq 0.17\times 0.04\times 0.2\simeq 1.4\cdot 10^{-3}$\,. 
Even with extra factors $\sim 0.1$ which may come from the
detailed calculations this is well within the 
modern experimental possibilities.

Finally, we mention a few possible signatures of SBNs  which can be used
for their experimental detection.
First of all, we remind the reader that according to the Dirac picture,
any real antibaryon should be interpreted as a hole in the otherwise filled
Dirac sea. Therefore, the nucleons from the positive-energy states of
the Fermi sea can make direct transitions to the vacant negative-energy states
of the Dirac sea. These super-transitions will be accompanied by the emission
of a single pion or kaon depending on the nature of the traped antibaryon.
The energy of such a super-transition is fixed by the discrete levels
of the initial and final baryons and according to our calculations
should be of about 1 GeV. This 1-pion or 1-kaon annihilation
is a unique feature of finite nuclear systems.
In vacuum such transitions are forbidden by the energy-momentum conservation.
Therefore, the observation of a line in the pion or kaon spectrum
at energies between 1 and 2 GeV would be a clear signal of the deep
antibaryon states in nuclei. One can also look for narrow
photon lines with energies in the range from 40 to 200 MeV
corresponding to the transitions of nucleons and antibaryons between their
respective levels. It is interesting to note that
these signals will survive even if due to the lack of time the nucleus
does not fully rearrange to a new structure.
 Another strong signal may come from the response of the nuclear
 remnant to the annihilation
of the antibaryon in the deeply bound state. Since the remnant nucleus
will initially be in a highly compressed state, it will expand and
eventually break up into fragments. Therefore, the annihilation process
will lead to rather cold multifragmentation with large collective flow
of fragments.  Both proposed signatures require rather ordinary
measurements, which should be easy to perform with standard detectors.

\section{Discussion and conclusions}
Our main goal with this paper was to demonstrate that energetic
antiproton beams can be used to study new interesting phenomena in
nuclear physics. We discuss the possible existence of a completely new kind
of strongly interacting systems where both the nucleons and the
antinucleons coexist within the same volume and where annihilation is
suppressed due to the reduction of the available phase space. Such
systems are characterized by large binding energy and high nucleon density.
Certainly, antinucleons can be replaced by antihyperons or even by
antiquarks. We have presented the first self--consistent calculation of
a finite nuclear system containing one antiproton in a deeply bound
state. For this study we have used several versions of the  RMF model
which give excellent description of ordinary nuclei. The presence of an
antiproton in a light nucleus like $^8$Be or $^{16}$O changes
drastically the whole structure of the nucleus leading to a much more
dense and bound state. Even stronger effects are expected 
in the $^4$He nucleus. It is clear however that these structural changes
can occur only if the life time of the antibaryons in the nuclear
interior is long enough.

One should bear in mind that originally the RMF model was
formulated within the Hartree and no--sea approximations. 
Implementing the Dirac
sea  may require serious revision of the model and inclusion of
additional terms. Hartree calculations including the Dirac
sea~\cite{Mao99} and Hartree--Fock
calculations~\cite{RMF,HFC2} including exchange terms lead to 
smaller nucleon potentials in normal nuclei. Shallower potentials
will produce smaller attraction for antinucleons, but the
qualitative effect 
that the presence of antiprotons reduces repulsion and enhances attraction
for nucleons will remain valid. We expect that the 
additional binding and compression of the
nucleus will appear even for an antinucleon potential as low as 200 MeV.

Since nucleon densities in the considered systems 
could reach $(2-3)\,\rho_0$\,, it becomes
questionable if the RMF model with nucleonic degrees of freedom is
applicable at all. Coupling constants of RMF models, obtained by the
fitting procedure, are predominantly constrained by observables at
saturation density. On the other hand, the nuclear systems studied
here  are sensitive to the equation of state far above the saturation
density. 
At such high densities nucleons might have to be substituted  by quark
degrees of freedom or modified by an admixture of them.  Recently, we have used an
extended version of the Nambu--Jona-Lasinio model to study bulk
properties  of systems composed of quarks and antiquarks~\cite{Mis99}.
It has been found that deeply bound states also appear in this model, but the
corresponding binding energies are considerably smaller by about a factor
of~3.

In summary, on the basis of the RMF model we have studied the structure
of nuclear systems containing a few real antibaryons.
We have demonstrated that the antibaryons act as strong attractors
for the nucleons leading to enhanced binding and compression of the
recipient nucleus. As our estimates show the life times of antibaryons
in the nuclear environment could be significantly enhanced due to the
reduction of the phase space available for annihilation.
Narrow peaks in the pion or kaon spectra at the energy around 1 GeV are
proposed as the most clear signature of deeply-bound antibaryon states
in nuclei.

\section*{Acknowledgements}
I.N.M. and L.M.S. acknowledge financial support from DAAD
and GSI, Germany. This work has been partially supported by the 
RFBR Grant No.~00--15--96590.

\end{document}